# Software Tools for Big Data Resources in Family Names Dictionaries

Adam Rambousek
Masaryk University, Faculty of Informatics
Botanicka 68a, Brno, Czech Republic
rambousek@fi.muni.cz

Harry Parkin
University of the West of England
Coldharbour Lane, Bristol BS16 1QY, United Kingdom
harry.parkin1@gmail.com

Ales Horak
Masaryk University, Faculty of Informatics
Botanicka 68a, Brno, Czech Republic
hales@fi.muni.cz

**Abstract:**

This paper describes the design and development of specific software tools used during the creation of Family Names in Britain and Ireland (FaNBI) research project, started by the University of the West of England in 2010 and finished successfully in 2016. First, the overview of the project and methodology is provided. Next section contains the description of dictionary management tools and software tools to combine input data resources.

## Family Names in Britain and Ireland Project Overview

FaNBI is an AHRC-funded research project, the aim of which is to research the origins of family names found throughout Britain and Ireland, and to produce a born-digital dictionary which is much more reliable and complete than those which have been published before. The project is based at the Bristol Centre for Linguistics at the University of the West of England, Bristol, with technical collaboration of members of the Faculty of Informatics at Masaryk University, Brno, in the Czech Republic. Linguistic and historical research, including the composition of dictionary entries, is carried out by the project staff based in Bristol, while members of Masaryk University's Faculty of Informatics are responsible for the creation, structuring, and maintenance of the database and data which are used by the team in Bristol.

Family name research has received relatively little scholarly attention in Britain, with place-names appearing to be much more popular as an onomastic subject. While a number of regional studies of English surname history have been produced, (see, for example, the English Surnames Series: Redmonds 1973; McKinley 1975, 1977, 1981, 1988; Postles 1995, 1998), and there has been much genealogical research into the history of individual family names (see the one-name studies carried out by members of the Guild of One-Name

Studies[1]), there are very few dictionaries available which reliably explain the origins of the nation's current family names. In comparison, there are many reliable place-name dictionaries available, including the county volumes produced by the English Place-Name Society. With a growing public interest in genealogical research, and an increasing availability of online resources for investigating family history, information on family name etymology is likely to be of great interest to many different people. Unfortunately, much of the previous research on surname origins is unreliable, and can easily mislead. The aim of the FaNBI project was to provide a new dictionary of surnames in Britain and Ireland, analysing new data and adopting new methodological approaches in order to explain the origins of family names which have not been studied previously, and to correct many of the mistakes which have been made before.

The earliest known work in which the origins of British surnames are discussed is Camden's (1605) *Remains Concerning Britain*. He dedicates one chapter to the subject, in what is mainly a discursive account of the history and development of surnames in Britain, but there are also some alphabetical lists of different types of surnames, making Camden's work the first which resembles a dictionary of British surnames, though it is far from extensive. The first dictionary to attempt an explanation for a significant proportion of British surnames was not produced until the early 20th century, when Bardsley (1901) wrote *A Dictionary of English and Welsh Surnames*. This work supplied examples of early bearers alongside surname explanations, thus highlighting the importance of linguistic research in determining surname origins.

Perhaps the most well-known dictionary is Reaney's (1958) *Dictionary of British Surnames*, later published as *A Dictionary of English Surnames* (Reaney and Wilson 1991) in a 3rd edition. Reaney's dictionary is a great achievement, especially given that the substantial early bearer evidence he presented was collected from such a wide range of sources, well before the widespread accessibility of historical records provided by the internet. However, his work is not without error. A number of his explanations have been shown to be incorrect, and in many cases this is due to Reaney's tendency to disregard a surname's distribution in the investigation of its origin, even though Guppy (1890) had already shown that there was often a clear link between the present-day distribution of a surname and its place of origin.

One key aspect of the FaNBI methodology, which sets it apart from previous works, is the systematic consideration of distribution for every dictionary entry. While this would have been difficult in the past, the relatively recent creation of the *British 19th Century Surnames Atlas* (Archer 2011) allows for the frequency and distribution of every surname in the 1881 census to be viewed clearly. This information can provide an important clue as to a surname's origin, especially for locative names which derive from particular places and are often still found close to those places today. In addition, surname distribution can be important for genealogy, guiding family historians to the parts of the country where they might benefit from carrying out further research. Given the value of geographical information, 1881 census distributions are concisely summarized and presented in written form in almost every FaNBI entry. Surname frequencies in 1881 and 2011 are also provided.

Another advantage of FaNBI is its extensive list of surname entries, providing explanations for a much greater number of names than attempted by previous works. The headword list is made up of almost all surnames with 100 or more bearers in the 2011 UK census, meaning that names with various ethnic and cultural origins are included. More entries were then added to the dictionary, so that names of particular philological interest, and names which have appeared in previous dictionaries, could also be included. All of this means that FaNBI contains over 45,000 surnames, approximately 20,000 of which are treated as 'main entries' with early bearer information and etymological explanations, while over 25,000 are spelling variants of the 'main entries'. As would be expected, the majority of the surnames in FaNBI

1 http://one-name.org/

are of English, Scottish, Welsh or Irish origin, though there are many surnames with other linguistic origins which have been established in Britain and Ireland for centuries, such as French, Huguenot, and German names. There are also approximately 3,800 entries in FaNBI which are classified as 'recent immigrant names', meaning they only became established in Britain and Ireland after World War II. These include names from China and the Indian subcontinent. These types of names rarely appear in previous surname dictionaries of British and Irish surnames, but are included in FaNBI in order to show the true multicultural nature of the current name stock.

As mentioned previously, the aim of the FaNBI project has been to produce the most reliable British and Irish family name dictionary available. This has been made possible by the increased availability of historical records in digital form, from which early family name spellings have been extracted and used as evidence for family name etymology and development. Some of the most important sources of early bearers, which have not been used in this way before, include the poll tax returns of 1377, 1379, and 1381 (provided by Fenwick 1998, 2001, 2005), Parish Registers (extracted from the International Genealogical Index (IGI)), PROB 11 probate returns, and Irish Fiants, some of which are discussed further later in the paper. For a detailed account of the FaNBI methodology, see Hanks, Coates, and McClure (2012).

## Editing and Management Application

A custom web application was developed for the editing of the dictionary content and management of the work progress in the FaNBI project. This application is based on the Dictionary Editor and Browser (DEB) platform developed by the Natural Language Processing Centre, Masaryk University. A predecessor of the generalized dictionary platform was the VisDic tool (Horak, Smrz, 2004), which was used by teams within the EuroWordNet project. Since 2005, DEB has been employed in more than 20 international research projects. Examples of applications based on the DEB platform include the Czech Lexical Database (see Horák, Rambousek, 2013) with detailed information on more than 213,000 Czech words, or the complex lexical database, Cornetto, combining the Dutch wordnet, an ontology, and an elaborate lexicon, see Horák, Vossen, Rambousek (2008), or DEBVisDic (Horak, Rambousek, 2007), the wordnet editor used for development of many national wordnets around the world. Current ongoing projects include the Pattern Dictionary of English Verbs tightly interlinked with corpus evidence, see Maarouf et al. (2014), and a compilation of the Dictionary of the Czech Sign Language with extensive use of multimedia recordings to present the signs visually, see Rambousek, Horak (2015).

Apart from the research team at UWE, many consultants from various institutions all over the world were involved in the FaNBI project. For this reason, the editing client software was designed as a web application for multi-platform use (see Figure 1). The application is implemented in standard HTML and JavaScript for the best compatibility with all modern web browsers.

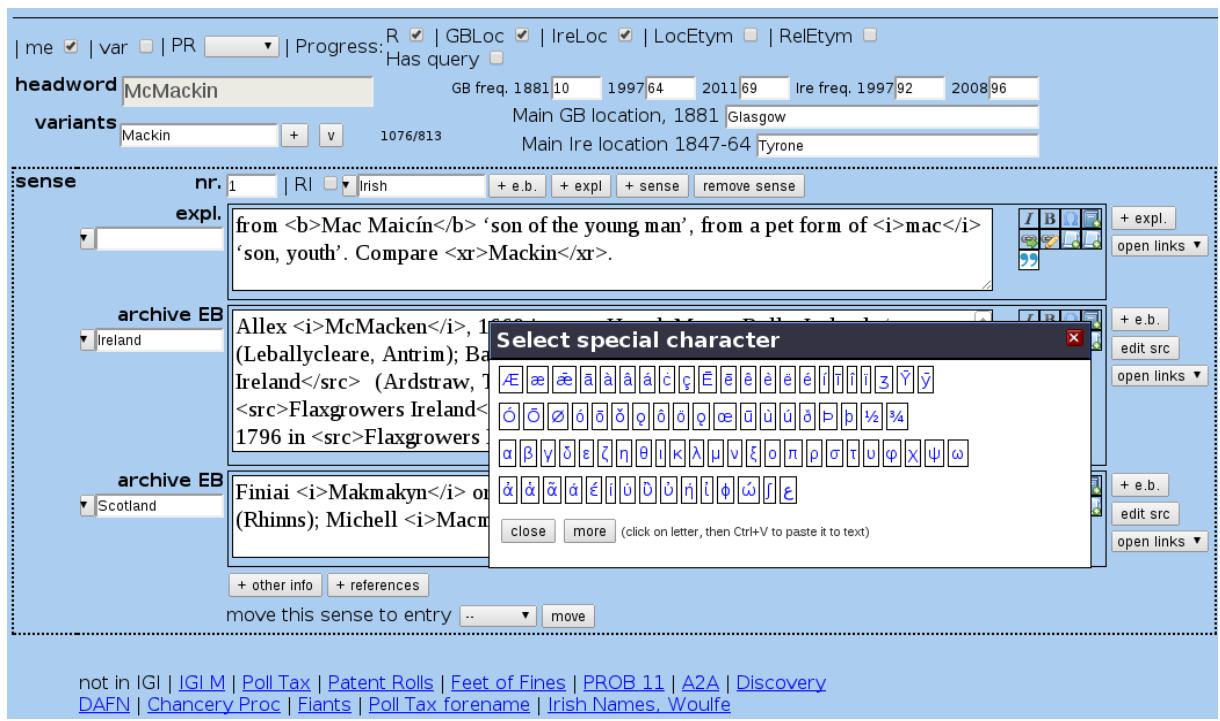

Figure 1: FaNBI – editing entry, with virtual keyboard.

The FaNBI database enables each family name to have several variant spellings or derived names. In order to keep all the information in one place, the names are grouped into "clusters" of related names. When the author wants to edit a family name, a complete cluster is opened in the editing application. It is possible to quickly move information (e.g. explanation) from one entry to another, add another name to the cluster, or select the main name for the cluster. The application updates cross-references automatically in such cases.

Each entry contains statistical information about the family name – the frequency of the name according to predefined records (data from 1881, 1997, and 2013 are included for Great Britain, and from 1997 and 2008 for Ireland) and the most prominent location where the family name is used. The system also allows to keep track of the work progress and share comments or requests with other team members.

Each dictionary entry may be composed of several senses describing the name origin, usually specified by the language or culture of origin (e.g. Welsh, Jewish, Arabic...). Each sense may contain several explanations describing the name origin from various point of views, a list of evidence of "early bearers" from historical sources (when possible, at least one person for each century and place where the name was frequent enough), and references and links to other resources. Whenever appropriate, the application provides templates for frequently repeated parts of the description. Within the explanations of family names origin, the authors need to enter characters from foreign alphabets, for which purpose a virtual keyboard is provided to select characters quickly. A convenient lookup for the family name in applied resources, both stored in the FaNBI DEB database or external web services, is also available.

For the sake of the long-term project management, the application offers specific tools to the lead project editors (see Figure 2). Statistics of the work done by the authors are generated weekly. The editors may also see a progress report of the whole project, either for the whole database, or a selected range. The report shows the current status of entries (e.g. how many of them are finished, not yet edited, etc.) and the number of entries in various categories

(e.g. main entry, variant entry). Any comments or requests blocking the entries are reported, too.

Figure 2: Management tool listing entries and their current state

# List of Names and Frequencies

The frequency list of each family name is the cornerstone of the FaNBI project work. It is not only the list of entries to edit, but the frequency also decides which names will be edited in each phase of the project. In the first phase, all names with more than 100 bearers were selected. The list was extended to all names with more than 20 bearers in the second phase.

At the beginning of the project, two lists were used – 1881 census report (National Archives) and 1997 statistical data (Hanks, Coates, 2012). However, both lists had to be preprocessed and filtered, since they contained a lot of noise and errors (for example, spelling errors or invalid characters). Another issue with the lists provided was that all the names were written in uppercase. A straightforward solution is to leave the first letter of each word uppercase and the rest in lowercase, however, this is not true for all names. For example, Scottish and Irish names like O'Brian or McGaffin had to be considered. This type of names also produced the issue with various written spellings used. For the Mc- names, three different variant spellings were present – Mac-, Mc- and M'-.

Similarly for O'- names, various apostrophe characters were used and sometimes the name was written without the apostrophe. It was decided to include only the spellings Mc- and O'- into the dictionary, and redirect readers searching for other variants to the correct dictionary entry.

To make the matter even more complicated, some family names starting with the string "Mac-" are separate names and not the variant spellings of "Mc-", for example Mach or Mackarel. To solve this issue, the list was edited in two steps. In the first step, variant spellings were detected and uncertain samples were reported. In the next step, the proposed changes were approved by the lexicographers. In case of variant spellings, the frequencies had to be summed for all the forms.

The 1881 census list was edited with this method, and then the method was used for updating the other lists. The lexicographers' approval was not needed anymore, since the 1881 list was included in the cleaning tool to decide the correct spellings and variant combination. Finally, only names with frequency of at least 20 were included. During the cleaning and combining of the 1881 census list, the number of records was reduced from 469,356 to 373,319 records.

With new reports from recent years, an issue linked to growing immigration was discovered – both masculine and feminine forms of the names appeared for languages where these forms differ (for example, Polish or Czech). It was decided to keep only the masculine form of the family name, and the method for frequency inclusion was updated. Feminine forms are detected by a known suffix and if the masculine form is present in the database, the frequency numbers are combined.

## Combining Resources

With the aim to include as much historical evidence as possible, various existing databases are used to search for the records of the family names. A selection of records is available as a webservice from The National Archive[2] , however it was needed to clean or preprocess the resources.

The International Genealogical Index (IGI) compiled by The Church of Jesus Christ of Latter-day Saints represents a very valuable resource for family names studies. IGI contains worldwide records extracted from the parish archives and similar sources, or submitted by the members of the Church. IGI records are published on the FamilySearch website[3], however, the website does not provided access to the complete collection and records may contain errors or inconsistencies. For the purpose of the FaNBI project, the original database records for the Great Britain were obtained. The database was transcribed from the parish archives by volunteers over the course of several decades. Because of many reasons (for example, unreadable books, different spellings by each transcriber, spelling mistakes etc.), the database had to be cleaned up before it could be included in the FaNBI research. Sometimes, several volunteers independently transcribed the same parish records, so the duplicate data had to be detected. The following list sums up the process of cleaning and deduplicating the IGI database.

- Original database contained 188,043,185 records. Each record contains information about the event type (birth, christening, marriage, or death), first name, surname, date, location (county, town/place name, sometimes the exact parish), and the role of the person (e.g. for marriage the bride, groom, or their parents).

- Obvious mistakes were deleted, for example records claiming that the English cities are in France.

- Names of the counties were standardized from variant spellings and abbreviations.

- For each county, a list of place names was extracted. These lists were distributed amongst the volunteers from the Guild of One-Name Studies. Volunteers checked if the place name on the list belongs to the given county, or they provided a correct spelling. As a result of this process, a standardized list of place names was created and the database records were fixed. The records with incorrect information about a place name were deleted.

- In the next step, duplicate records were deleted. Since the main aim of this process for the FaNBI research was not to build complete and perfect database, but provide reliable evidence, it was possible to delete not just exact duplicates, but also suspect duplicates. The rules for duplicate detection were based on the following information from the records: the first name, surname, date, town, county, and event type. Records were flagged as duplicate when all information was identical, except one of the following fields was different: first name, town, county, or event type.

- At the end of the process, IGI database contained 72,187,630 records. For a sample of the original IGI record and the converted form to include as the historical evidence see Table 1. The *original record* contains information about one event with items separated by the '|' character. This particular record talks about christening of a person, whose father was named John Darter, and which took place in Bletsoe in the county of Bedfordshire. The *converted record* shows the father's full name (with markup for surname), with year, source and place references (Beds is a FaNBI shortcut for Bedfordshire).

Subsequently, the database was used to automatically add historical evidence to the FaNBI dictionary. For each family name, IGI records were extracted for each century and most prominent county, formatted according to the reference templates and saved in the entry. 40,274 family names entries were automatically enhanced with the IGI evidence. Apart from the enhancement of the dictionary, the processed IGI database is regularly consulted by the researchers as a valuable resource.

Table 1: Original record from the IGI database and the form included into FaNBI.

| |
|---|
| *Original record* (batch identification, event date, event place, event type, year, first name, surname, role, gender): <br><br> Bletsoe, Bedford, England\|05 Sep 1629\|Bletsoe, Bedford, England\|Christening\|1629\| John\|Darter\|Principal's Father\|Male |
| *Converted record:* <br><br> John <sn>Darter</sn>, 1629 in <src>IGI</src> (Bletsoe, Beds) |

Another archive resource that required preprocessing consisted of the three volumes of *The Irish Fiants of the Tudor sovereigns during the reigns of Henry VIII, Edward VI, Philip & Mary, and Elizabeth I* (Nicholls, 1994). The Fiants contain court warrants and their texts are available in the electronic format obtained by digitization and OCR of the original papers. Each record is clearly marked in the text and thanks to the official language, it is possible to detect persons' names, occupations, or residence. Within the FaNBI project, the Word documents, as obtained from the OCR recognition process, were converted to the XML format. Each court record was converted into a separate XML entry with enhanced metadata. For example, the date of the record was converted from the regnal years system into the standard Gregorian calendar years.

The annotated XML documents were then processed by a developed information extraction tool. The tool standardized common OCR misspellings and detected frequently repeating text patterns in the warrant texts. The list of place names (obtained in the IGI database cleanup) was used to detect town names and match them with the correct county. Where available, also the persons' occupations were tagged in the record. Finally, all the information was formatted according to the FaNBI reference templates and it is available for reference in appropriate entries. For a sample of the conversion from Fiants to FaNBI, see Table 2. The *original record* numbered 1431 represents a pardon to a specific person together with the place and date of the event (in the regnal years format) and his occupation. The *converted record* in FaNBI extracts the same information and format as used in Table 1 with IGI data, only the Fiants record number is added as '$1431' for reference here.

Table 2: Original Fiants record and the converted form to be included in FaNBI.

| |
|---|
| *Original record:* |
| 1431. Pardon to Thomas Dowdall, of Dermondston, county Dublin, husbandman.— 2 November, xi. |
| *Converted record:* |
| Thomas <sn>Dowdall</sn>, 1569 in <src>Fiants Eliz</src> $1431 (Dermondston, co. Dublin) |

## Preparing Data for the Publisher

The resulting dictionary of family names was published by the Oxford University Press (OUP), in November 2016. During the development of the tools, the XML document format for the publication was discussed and developed in coordination with OUP. Thanks to the design of the DEB platform output formatting, it was possible to test several prototypes before agreeing on the final delivery format. In case of updates (for example, enhanced dictionary data, or fixed spelling mistakes), the updated data were quickly prepared in the right format for publication. The dictionary data were validated before sending to the publisher. For example, the correct publishing layout or completeness of the whole dictionary data was checked in this phase.

## From FaNBI to DAFN2

The presented software tools were re-implemented for the currently ongoing editing of the Dictionary of American Family Names 2nd edition (DAFN2), with the expected size of 80,000 entries, lead by the chief editor Patrick Hanks and to be published by the Oxford University press. The verified methodology and tools from the FaNBI project were adapted for the DAFN2 dictionary. The editing and management application is customized to meet different requirements. Methods and tools developed for tidying the names frequency lists were reused, because the DAFN2 project needs to solve similar issues. Thanks to the DEB API interface, the applications are able to easily share also the data, thus extending the research possibilities. And finally, since the publisher for both dictionaries is OUP, the format and tools implemented for the publication of results may be reused.

# Conclusions

We have presented the methodology and software tools used to build two large projects aimed at the family names research - the Family Names in Britain and Ireland (FaNBI) project and the Dictionary of American Family Names 2nd edition (DAFN2). Both projects combine numerous archive and modern resources and the dictionary management tools provide effective ways to extract important information from diverse sources and formats.

The FaNBI project was already successfully finished and published in The Oxford Dictionary of Family Names in Britain and Ireland (2016).

Acknowledgements

This work has been partly supported by the Ministry of Education of CR within the LINDAT-Clarin project LM2015071. Family Names of the United Kingdom research was funded by the Arts & Humanities Research Council in projects AH/H018921/1 and AH/L007401/1.

# Notes on contributor

Adam Rambousek is a Research Assistant at Faculty of Informatics, Masaryk University, Brno, Czech Republic. His main interests are computational lexicography, corpus linguistics, ontologies and semantic networks.

Harry Parkin is a Research Associate at the University of the West of England, Bristol, United Kingdom. His research focuses on dialectology and onomastics and he is member of the Family Names in Britain and Ireland research team.

Ales Horak is an Associate Professor of Informatics at Masaryk University, Brno, Czech Republic. His research concentrates on natural language processing, knowledge representation and reasoning, e-lexicography and corpus linguistics.

**Correspondence to**

Adam Rambousek
Masaryk University, Faculty of Informatics
Botanicka 68a, Brno, Czech Republic
rambousek@fi.muni.cz